\documentclass[amsmath,twocolumn, secnumarabic,amssymb,nobibnotes,aps,showpacs,nofootinbib,noeprint,superscriptaddress]{revtex4-1}
\usepackage[english]{babel}
\usepackage{graphicx, amsfonts, amssymb}
\usepackage{hyperref}
\usepackage{colortbl}
\usepackage[sort&compress,capitalize]{cleveref}
\usepackage{units}
\newcommand{\dd}{\text{d}}
\newcommand{\kB}{k_\textrm{B}}
\newcommand{\ld}{\lambda_\textrm{D}}
\newcommand{\blue}[1]{\textcolor{blue}{#1}}
\definecolor{purple}{rgb}{0.6,0.0,.5}

\DeclareMathOperator\arctanh{arctanh}
\DeclareMathOperator\arcsinh{arcsinh}
\DeclareMathAlphabet{\mathbb}{U}{bbold}{m}{n}
\begin{document}
\title{
Reversible heat production during electric double layer buildup depends sensitively on the electrolyte and 
its reservoir
}

\author{Fabian Glatzel}
\email{fabian.glatzel@physik.uni-freiburg.de}
\affiliation{Institute of Physics, University of Freiburg, Hermann-Herder-Str. 3, 79104 Freiburg, Germany}

\author{Mathijs Janssen}
\email{mathijsj@uio.no}
\affiliation{Mechanics Division, Department of Mathematics, University of Oslo, 0316 Oslo, Norway}

\author{Andreas H\"{a}rtel}
\email{andreas.haertel@physik.uni-freiburg.de}
\affiliation{Institute of Physics, University of Freiburg, Hermann-Herder-Str. 3, 79104 Freiburg, Germany}

\date{\today}

\begin{abstract}
Several modern technologies for energy storage and conversion are based on the screening of 
electric charge on the surface of porous electrodes by ions in an adjacent electrolyte. 
This so-called electric double layer (EDL) exhibits an intricate interplay with the electrolyte's 
temperature that was the focus of several recent studies. 
In one of them, Janssen \textit{et al.} 
[\href{https://journals.aps.org/prl/abstract/10.1103/PhysRevLett.119.166002}{Phys. Rev. Lett. \textbf{119}, 166002 (2017)}] 
experimentally determined the ratio $\mathcal{Q}_\text{rev}/W_\text{el}$ of reversible heat flowing 
into a supercapacitor during an isothermal charging process and the electric work applied therein. 
To rationalize that data, here, we determine $\mathcal{Q}_\text{rev}/W_\text{el}$ within different 
models of the EDL using theoretical approaches like density functional theory (DFT) as well as molecular 
dynamics simulations. Applying mainly the restricted primitive model, we find quantitative support for 
a speculation of Janssen \textit{et al.} that steric ion interactions are key to the ratio 
$\mathcal{Q}_\text{rev}/W_\text{el}$. 
Here, we identified
the entropic contribution of certain DFT functionals, 
which grants direct access to the reversible heat. 
We further demonstrate how $\mathcal{Q}_\text{rev}/W_\text{el}$ changes when calculated in 
different thermodynamic ensembles and processes. We show that the experiments of 
Janssen \textit{et al.} are explained best by a charging process at fixed bulk density, 
or in a ``semi-canonical'' system. 
Finally, we find that 
$\mathcal{Q}_\text{rev}/W_\text{el}$ 
significantly depends on parameters 
as
pore and ion size,
salt concentration, and
valencies of the cat- and anions of the 
electrolyte. Our findings can guide further heat production measurements and can be applied in studies 
on, for instance, nervous conduction, where reversible heat is a key element. 
\end{abstract}

\maketitle

\section{Introduction}
In a recent experiment \cite{janssen_prl119_2017}, the reversible heat flowing into and the 
electric work applied to a supercapacitor during isothermal charging were measured. 
This experiment is one of several recent studies, both 
experimental \cite{schiffer_jps160_2006,janssen_prl119_2017,lindner2020entropy,rodenburg_MSc_2020} 
and theoretical \cite{dEntremont_jps246_2014,*dEntremont_jps273_2015,kumar_arxiv_2015,janssen_prl118_2017,cruz2018electrical,deLichtervelde_pre101_2020,alizadeh_ep41_2020}, 
on the intricate interplay 
between the electrolyte's temperature and the properties of the electric double layer (EDL). 
Until now, however, no comparison has been made between the experimental findings 
of Ref.~\cite{janssen_prl119_2017} and theoretical predictions from sophisticated EDL models.

\textsc{Helmholtz} proposed the EDL to be a system where two layers of opposite 
charges are facing and, thus, screening each other \cite{helmholtz_ap165_1853}. Usually, systems are considered where mobile ions physically 
screen electric charge, for instance, on a solid electrode's surface \cite{conway_book_1999}, 
on colloidal particles \cite{verwey1948theory,derjaguin_pss43_1993}, 
in (biological) ion channels of the plasma membrane \cite{roth_bj94_2008,Roth2014}, 
and near macromolecules such as DNA \cite{kornyshev_rmp79_2007}. 
The resulting diffuse double layer for 
point-like ions was first described by \textsc{Gouy} and \textsc{Chapman} around 1910 within \textsc{Poisson-Boltzmann} theory 
\cite{gouy_1910,chapman_1913,barrat_book_2003}, 
a framework even nowadays still applied frequently to study electrolyte systems. 
This simple picture of point charges is refined 
in more sophisticated models that account for finite ionic volume, where the latter can be important 
for the microscopic structure of EDLs in narrow geometries and crowded environments. For instance, 
the finite volume of ions can be crucial for the description of colloidal interactions, 
capacitances, and understanding certain aspects of screening 
\cite{hansen_arpc51_2000,haertel_jpcm29_2017,coupette_prl121_2018}. 

In modern technologies, EDLs also form the basis for the aforementioned supercapacitors, which can 
store much more energy than conventional capacitors and can deliver much higher power than batteries 
\cite{simon_nm7_2008,raghavendra_jes31_2020,simon2020perspectives}. 
For practical applications wherein these devices are charged and discharged, it is important to 
know how the electrolyte temperature can be kept low, because increased temperatures are 
the cause of faster degradation of components. 
Interestingly, these EDL systems can further be employed to desalinate solutes \cite{suss_ees8_2015} 
and to harvest energy, because concentration \cite{brogioli_prl103_2009} and temperature 
\cite{janssen_prl113_2014,haertel_ees8_2015,cruz2018electrical} can change their capacitance. 
Accordingly, vast amounts of studies on EDL systems exist and are still performed, but the 
interplay between ions and the electrolyte's temperature is still rather unexplored, 
despite being promising for optimization and new concepts. 

Measurements of the temperature of a supercapacitor in operation showed that it heated during 
charging and cooled during discharging \cite{schiffer_jps160_2006}, showing an overal trend to warm 
up during cycling. Such a warm up is expected as ionic currents in a resistive fluid dissipate 
\textsc{Joule} heat. 
The cooling, however, can be understood from an analogy to the adiabatic decompression of an ideal gas: 
During discharging, ions leave the EDL and their entropy increases. In an isolated supercapacitor 
this increase must be balanced by an entropy decrease of the electrolyte, accomplished through a lowering of the 
electrolyte's temperature. The opposite happens during charging and causes heating additional to \textsc{Joule} heat. 
Moving beyond the above ideal-gas analogy, a thermodynamic identity was derived for the temperature 
rise upon adiabatic EDL formation \cite{janssen_prl113_2014}. 
Predictions of this identity coincided with numerical solutions of the electrokinetic equations for a slow 
charging process \cite{janssen_prl118_2017}. 
In the latter nonequilibrium framework, reversible and irreversible heating are both captured by the 
heat production term $\vec{I}\cdot\vec{E}$ in the heat 
equation \cite{landau2013electrodynamics,dEntremont_jps246_2014,janssen_prl118_2017}, 
with $\vec{I}$ the ionic current density and $\vec{E}$ the local electric field. 
While the ionic current density aligns with the local electric field $\vec{I} \propto\vec{E}$ in 
bulk electrolytes, leading to a strictly positive \textsc{Joule}-heating term $\propto\vec{I}^2$, 
conversely $\vec{I}\cdot\vec{E}<0$ is possible in the EDL when the gradient in electrochemical potential 
anti-aligns with $\vec{E}$, leading to reversible 
local cooling \cite{dEntremont_jps246_2014,*dEntremont_jps273_2015,janssen_prl118_2017}. 
Similar cooling has been observed near an ion-exchange membrane \cite{porada2019electrostatic,biesheuvel2020physics}. 
As \textsc{Joule} heating is mainly a bulk phenomenon, while reversible heating happens 
only in the nanometer-wide EDL, a capacitor with a large surface-to-volume ratio is needed to notice 
appreciable reversible temperature variations. 
Advanced ``microcalorimetry'' measurements near flat electrodes, however, can detect 
much smaller temperature variations \cite{lindner2020entropy}. 

\begin{figure}
	\centering
 	\includegraphics[scale=1]{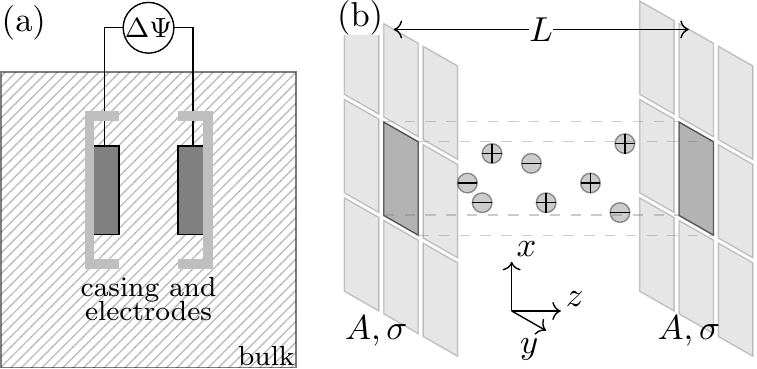}
	\caption{Sketch (a) of the experimental setup of Ref.~\cite{janssen_prl119_2017}. 
  In the experiments, two nanoporous carbon electrodes that carry charges $\pm Q$ due to a potential 
  difference $\Delta\Psi$ are in contact with an 
  aqueous sodium chloride solution and submerged in a thermostatic bath with respect to which its 
  temperature is measured. 
  Within our theoretical description, the pores of one electrode are modeled (b) by two planar walls of 
  area $A$ with equal surface charge density $\sigma$. The electrolyte between the walls is modeled by 
  mobile ionic charges. 
  }\label{pic:setup}
\end{figure}

With a setup as sketched in \cref{pic:setup}(a), the authors of Ref.~\cite{janssen_prl119_2017} 
studied the temperature 
of and charge on nanoporous carbon electrodes subject to a suddenly applied potential change. 
From the difference $\mathcal{Q}_\text{tot}-\mathcal{Q}_\text{irr}$ between the total and irreversible 
(\textsc{Joule}) heat, for which they had independent measurements, 
they determined the reversible heat $\mathcal{Q}_\text{rev}$, \textit{i.e.}, the temporal and 
spatial integral of the above heat production for slow charging (cf.~\cref{eq:QrevII}). 
Moreover, from the system's capacitance they determined the electric work $W_\text{el}$ during isothermal charging. 
With $\mathcal{Q}_\text{rev}$ and $W_\text{el}$ at hand, 
the authors of Ref.~\cite{janssen_prl119_2017} claimed experimental access 
to the ratio $\Delta\Omega_\text{ent}/\Delta\Omega$ (reproduced here in \cref{fig:heatWorkRatios}), 
with $\Delta\Omega$ the change of the total grand potential during charging and $\Delta\Omega_\text{ent}$ 
its entropic part. 
Their identification $\mathcal{Q}_\text{rev}/W_\text{el}=-\Delta\Omega_\text{ent}/\Delta\Omega$ relied on the identity 
$\mathcal{Q}_\text{rev} = -\Delta\Omega_\text{ent}$ (cf.~\cref{eq:qref_omega_ent}), proposed by 
\textsc{Overbeek} on thermodynamic grounds \cite{THEODOOR199061}, and on $\Delta\Omega = W_\text{el}$, 
which holds for isothermal charging.
The linear scaling of both $\mathcal{Q}_\text{rev}$ and $W_\text{el}$ with the surface area of the electrodes drops in their ratio, making $\mathcal{Q}_\text{rev}/W_\text{el}$ a quantity that can be conveniently compared with theoretical model predictions.
In fact, $\Delta\Omega_\text{ent}$ and $\Delta\Omega$ had been studied before within \textsc{Poisson-Boltzmann} 
theory \cite{THEODOOR199061} and extensions thereof accounting for 
finite-size ions \cite{kralj1996simple,biesheuvel2007counterion}. 
In particular, Ref.~\cite{biesheuvel2007counterion} found that the Carnahan Starling bulk chemical 
potential radically altered $\Delta\Omega_{\rm ent}/\Delta\Omega$ at large voltages: 
while \textsc{Poisson-Boltzmann} predicts $\Delta\Omega_{\rm ent}/\Delta\Omega\to1$, their more sophisticated 
theory suggested $\Delta\Omega_{\rm ent}/\Delta\Omega\to0$ instead. For this reason, 
the authors of Ref.~\cite{janssen_prl119_2017} speculated about the importance of ionic steric interactions to 
their measurement of $\Delta\Omega_{\rm ent}/\Delta\Omega\approx 0.25$ at $\Delta \Psi = \unit[1]{V}$.  

In this study we determine $\mathcal{Q}_\text{rev}/W_\text{el}$ within 
the restricted primitive model (RPM), where ions are described as charged hard spheres and the solvent is 
represented by an homogeneous background. 
The RPM is easy to simulate and is well described in classical density functional 
theory (DFT) \cite{haertel_jpcm27_2015,haertel_jpcm29_2017}, a version of the famous quantum 
DFT adopted to classical systems \cite{ebner_pra14_1976,evans_ap28_1979,hansen2013theory}. 
In our study, we apply several theoretical approaches of different sophistication: 
the (modified) \textsc{Poisson-Boltzmann} theory of Refs.~\cite{THEODOOR199061,kralj1996simple}, 
without and with a \textsc{Stern} layer,
a density functional theory with a very accurate description of the hard-sphere 
interaction \cite{haertel_jpcm29_2017}, 
and molecular dynamics (MD) simulations \cite{frenkel2001understanding}. 
For our DFT approaches we verify \textsc{Overbeek}'s identity 
$\mathcal{Q}_\text{rev} = -\Delta\Omega_\text{ent}$. 
Moreover, we discuss the importance of the choice of the thermodynamic process and the corresponding 
ensemble, which we demonstrate 
using the analytical \textsc{Gouy-Chapman} solution to the \textsc{Poisson-Boltzmann} 
equations. Finally, anticipating future experiments with other electrode--electrolyte combinations than 
used in Ref.~\cite{janssen_prl119_2017}, we study different pore sizes, ionic radii, 
valencies, and bulk ion concentrations. 

\section{Theory}

\subsection{Thermodynamics}

Thermodynamics allows us to draw general conclusions for our model system without 
using microscopic details. We consider the setup sketched in \cref{pic:setup}(a), 
where two porous electrodes are immersed in an electrolyte held at temperature $T$. 
The pores of each electrode have a certain fixed geometry during charging 
processes. 
Our system
contains an electrolyte with $N_+$ positive ions, $N_-$ negative ions, and 
$N_\text{s}$ neutral solvent particles. Conjugated to these particle numbers 
are the intensive chemical potentials $\mu_i$ with $i\in\{+,-,\text{s}\}$ 
that reflect the connection of the pore volume to an (infinitely) large reservoir that the 
electrodes are immersed in.

Upon connecting the electrodes to a battery that supplies a potential difference $\Delta\Psi$, 
the electrodes acquire electric surface charges $Q$ and $-Q$. 
The first law of thermodynamics for this system 
relates the change of internal energy $U$, heat $\mathcal{Q}$ transfered to the system, 
and thermodynamic work $W$ done to the system, and reads 
\begin{align}\label{eq:firstlaw}
 \dd U&= 	\delta \mathcal{Q} + \delta W \,.
\end{align}
The electric work performed during charging is given by 
\begin{align}\label{eq:work}
	W_\text{el} &= \int\limits_0^{Q}\Delta\Psi\ \dd Q'\,.
\end{align}
Using \textsc{Legendre} transforms, we obtain the free 
energy $F(T,Q,N_i)=U(S,Q,N_i)-TS$ and the grand potential 
$\Omega(T,Q,\mu_i)=F(T,Q,N_i)-\sum_i\mu_iN_i$, 
where the entropy $S$ enters. 

According to \cref{eq:firstlaw}, a process in, say, a grand canonical system 
wherein the surface charge on the positive electrode 
changes from $Q_{1}$ to $Q_{2}$ must cause a (reversible) heat flow into the capacitor that reads  
\begin{subequations}
\label{eq:defQrev}
\begin{align}
\label{eq:Qrev}
\mathcal{Q}_\text{rev}&= T\Delta S \, . 
\intertext{Here, the entropy difference $\Delta S$ is given by }
\Delta S &= 
-\left(\frac{\partial \Omega}{\partial T}\right)_{\mu_i,Q_2} 
+ \left(\frac{\partial \Omega}{\partial T}\right)_{\mu_i,Q_1} \, . 
\label{equ:entropy_difference}
\end{align}
\end{subequations}

Using \cref{eq:Qrev} and the Maxwell relation $(\partial S/\partial Q)_T=-(\partial\Delta\Psi/\partial T)_Q$, 
the heat during the isothermal charging process follows as 
(see also Eq.~(S7) of Ref.~\cite{janssen_prl119_2017}) 
\begin{subequations}\label{eq:Qrev1}
\begin{align} \label{eq:QrevI'}
\mathcal{Q}_\text{rev}^{\text{I}}&=
T\int_{Q_1}^{Q_2}\left(\frac{\partial S}{\partial Q'}\right)_{T,\mu_i}\dd Q' \\
&=-T\int_{Q_1}^{Q_2} \left(\frac{\partial \Delta\Psi}{\partial T}\right)_{Q',\mu_i}\dd Q'\, . \label{eq:QrevI'b}
\end{align}
\end{subequations}
In \cref{eq:QrevI'}, we introduce a superscript I for the reversible heat to distinguish it from the later 
result in \cref{eq:QrevII} which uses microscopic information and is valid only in a canonical system. 
Note that \cref{equ:entropy_difference,eq:QrevI',eq:QrevI'b} also hold in a canonical ensemble if the 
grand potential $\Omega$ is replaced by the free energy $F$ and the derivatives are taken at constant 
$N_i$ instead of $\mu_i$. 

For a general process where all the natural variables of a thermodynamic 
potential except for the surface charge are kept constant, the electric work corresponds to the 
change in the thermodynamic potential.
For instance, in a grand canonical system, the above charging process, where the chemical potentials 
are kept constant, results in the electric work 
\begin{equation}
W_\text{el} = \Delta \Omega\,.\label{equ:work_pot_equivalence}
\end{equation}

Likewise, the reversible heat can be expressed as the entropic contribution 
$-TS$ to the thermodynamic potential. 
Here, we already mention that both $F$ and $\Omega$ contain this term $-TS$. Thus, knowing 
expressions of the thermodynamic potentials and being able to identify the contribution $-TS$ 
would allow to directly read off the change of entropy and, hence, the reversible heat. 
Later we will see that this becomes handy in the framework of DFT.
From hereon, for convenience, we will use the uncharged electrodes as the reference state and, hence, the change in the grand potential $\Delta\Omega$ and its entropic contribution $\Delta\Omega_\text{ent}$ vanishes for 
$\Delta\Psi=0$. 
Regarding a grand canonical system, for example, this means that the reversible heat during 
charging satisfies 
\begin{equation}\label{eq:qref_omega_ent}
\mathcal{Q}_\text{rev} = -\Delta\Omega_\text{ent}\,.
\end{equation} 
Again, \cref{equ:work_pot_equivalence,eq:qref_omega_ent} also hold for a canonical system 
if $\Omega$ is replaced by $F$. 

\subsection{Microscopic model setup}
\label{sec:model}

To determine thermodynamic potentials and related state functions 
for the setup sketched in \cref{pic:setup}, we have to model the microscopic details of the 
capacitor system. 
To capture the essential physics of EDLs in nanopores, we 
model the pores of each electrode by two parallel planar walls of surface area $A$ 
and separation (pore size) $L$, as sketched in \cref{pic:setup}(b); note that for our MD simulations we used oppositely charged walls as explained in \cref{sec:mdSimulations,sec:simulationdetails}. 
Both walls combined carry the total charge $\pm Q$ of the respective electrodes, 
leading to a surface charge density $e\sigma=Q/(2A)$ with the proton charge $e$. 

To benefit from symmetries, we consider 
each pore wall stretching infinitely in the $(x,y)$ plane 
of a Cartesian coordinate system such that edge effects are suppressed. 
The pore walls are positioned at $z=0$ and $z=L$. 
In this setting, the EDLs at the left and right wall generally overlap.
However, if $L$ is sufficiently large, both EDLs can be considered independent, a situation we call 
free of overlap. In this case, a variation of $L$ does not affect the EDL. 
If the state is free of overlap and ions are symmetric, 
the study of only one EDL at one wall is sufficient, because the EDLs at all other 
walls of both electrodes will follow from symmetries. 

In this work, we mainly focus on one electrode (two walls) and define the pore volume $A L$ of one of the 
electrodes to contain $N_+$ positive ions, $N_-$ negative ions and $N_\text{s}$ neutral solvent particles. 
The ions of the electrolyte have valencies $z_{i}$ that define the number of positive 
unit charges $e$ per ion. Unless stated otherwise we consider $z_{\pm}=\pm 1$. 
In cases where we do not explicitly account for the volume of solvent particles we 
set $N_\text{s}=0$. While we generally assume solvent particles to have $z_{\text{s}}=0$, 
if explicitly present at all, the dielectric nature of the solvent is always accounted for in 
a dielectric background via a relative permittivity $\varepsilon_r$. 

The experiments of our interest dealt with porous carbon electrodes 
and aqueous sodium chloride \cite{janssen_prl119_2017}. 
While we usually use states free of overlap by setting $L$ to large values, 
choosing $L=\unit[1.6]{nm}$ would result in the same ratio of pore volume to 
electrode surface area as in the experiments. 
To account for the steric effects of the ions, we adopt a restricted primitive model (RPM) 
and describe the sodium and chloride ions as charged hard spheres. As diameter we choose $d=d_{\pm}=0.68$ nm, 
a value determined in scattering measurements and approximating their effective size in water 
\cite{doi:10.1021/j150579a011,haertel_jpcm29_2017}. 
We also study the solvent primitive model (SPM), 
an extension of the RPM where solvent particles are added as uncharged hard spheres 
of \mbox{$d_\text{s}=0.3$ nm} such that the total volume fraction is $46.8\%$ (corresponding to 
pure water in our model). 

Apart from our MD simulations, where particles are considered explicitly, our other theoretical treatments handle 
particle densities $\rho_i(\vec{r})$ of a particle species $i\in\{+,-,\text{s}\}$, 
\textit{i.e.}, the number of particles at a position $\vec{r}$ averaged over all 
states in an ensemble. 
Due to the infinite extension of the pore walls, the number densities 
$\rho_{i}(\vec{r})$ depend only on $z$. 
We further denote the respective bulk densities by $\bar{\rho}_i$. 
They have to satisfy the condition $0=\sum_iz_i\bar{\rho}_i$ to ensure electroneutrality in the bulk. 

We construct the local unit charge density as
\begin{align}\label{eq:chargedensity}
	q(z) &= \sigma [\delta(z) + \delta(z-L)]+\sum\limits_{i\in\{+,-\}} z_{i} \rho_i(z)\,, 
\end{align}
using \textsc{Dirac} $\delta$-distributions $\delta(z)$. 
The \textsc{Poisson} equation now relates $q(z)$ to the electrostatic potential $\psi(z)$ through
\begin{align}\label{eq:poisson}
	\varepsilon_{0}\varepsilon_{r} \partial_{z}^2 \psi(z) = -e q(z)\,,
\end{align}
with $\varepsilon_0$ the dielectric permittivity of the vacuum. 
We will frequently use the dimensionless potential $\phi(z) =e\psi(z)/\kB T$, 
with $\kB$ the \textsc{Boltzmann} constant. 
Using capital letters, we denote the electrode potential by $\Psi=\psi(z=0)$ and $\Phi=\phi(z=0)$. 
As sketched in \cref{pic:setup}(b), we set $\psi(0)=\psi(L)$. 

To come to a closed set of equations, \cref{eq:chargedensity,eq:poisson} need to be supplemented 
with an expression for $\rho_{\pm}(z)$ in terms of $\psi(z)$. 
Accordingly, in the following we present different theoretical approaches, formulated 
within the framework of classical density functional theory (DFT). 

\subsection{EDL modeling within classical density functional theory}\label{sec:DFT}

The central quantity in DFT is the grand potential $\Omega[\{\rho_i\}]$, a functional  
of the particle densities $\rho_i$ in the system. 
While the grand potential functional also depends on $T$ and $\mu_i$, 
for readability we omit these dependencies in our notation. 
As common, we split up the grand potential functional into 
\begin{align}\label{eq:OmegaDFT}
\Omega[\{\rho_i\}] &= \mathcal{F}_\text{id}[\{\rho_i\}] + \mathcal{F}_\text{exc}[\{\rho_i\}]  \nonumber\\
&\phantom{=} + A\sum_i \int \rho_i\left(z\right)\left[V_\text{ext}\left(z\right)-\mu_i\right]\dd z\,,
\end{align}
with the 
intrinsic free energy functional $\mathcal{F}_\text{id}[\{\rho_i\}]$ of an ideal gas \cite{hansen2013theory}, 
an excess free energy functional $\mathcal{F}_\text{exc}[\{\rho_i\}]$ that adds contributions due to 
pair potentials, and a contribution from an external potential $V_\text{ext}\left(\vec{r}\,\right)$ 
and chemical potentials (the latter enter in the \textsc{Legendre} transform between $F$ and $\Omega$). 
Importantly, the grand potential functional is minimal for the correct (physical) equilibrium 
particle densities of the system and its value then equals the value of the actual (thermodynamic) 
grand potential \cite{PhysRev.137.A1441}. This property allows to determine equilibrium 
density profiles by minimizing a given functional. 

While the ideal free energy functional is known exactly, exact excess free energy functionals 
are only known in a few cases. Nevertheless, 
many approximations have been tested for specific problems. 
In this work, we employ three 
well-established approximations to $\mathcal{F}_\text{exc}$; one to 
describe point-charge particles and two to describe particles that additionally occupy volume in space. 
For the point charges, we use a 
mean-field \textsc{Coulomb} functional (cf. \cite{haertel_jpcm29_2017}) that reads 
\begin{align}
\mathcal{F}_\text{C} &= \frac{eA}{2}\int q(z)\psi(z) \dd z\,. 
\end{align}
We refer to this simplest choice by PB, because the {\sc Euler-Lagrange} equations of this functional together with $\mathcal{F}_\text{id}$
yield the well-known {\sc Poisson-Boltzmann} equation \cite{barrat_book_2003}. 
For the next approach, 
we extend the above mean-field \textsc{Coulomb} functional by 
an excess lattice-gas functional $\mathcal{F}_\text{lg}$ that treats the occupied volume of the 
particles effectively via a maximum local number of allowed particles 
\cite{borukhov_prl79_1997}. 
The respective free energy functional apart from $\mathcal{F}_\text{C}$ reads 
\begin{align}
\mathcal{F}_\text{id} + \mathcal{F}_\text{lg} 
= k_\text{B}T A \int \Bigg( &
  \sum_{i\in\{+,-\}}\rho_{i}(z) \ln\left(\tfrac{\rho_{i}(z)}{\rho_\text{vac}(z)}\right) \notag \\
  & 
    - \rho_\text{M}\ln\left(\tfrac{\rho_\text{M}}{\rho_\text{vac}(z)}\right)
\Bigg) \dd z \,, \label{eq:functional_mPB}
\end{align}
where the density of lattice vacancies is $\rho_\text{vac}(z)=\rho_\text{M}-\rho_{+}(z)-\rho_{-}(z)$ 
with $\rho_\text{M}$ defining the highest local concentration or, in other words, the number density 
of accessible lattice sites. The latter is determined from assuming random close packing of hard spheres, 
resulting in $\rho_\text{M}d^3\pi/6=0.634$ \cite{song_nature453_2008}. 
We refer to this approach by mPB, because the functional in \cref{eq:functional_mPB} together with $\mathcal{F}_\text{C}$
yields the modified {\sc Poisson-Boltzmann} equation \cite{kralj1996simple,borukhov_prl79_1997} 
with $\beta\mu_\pm=\ln(\bar{\rho}_\pm/(\rho_\text{M}-2\bar{\rho}_\pm))$. 
Finally, we construct a functional for the RPM and SPM by extending the excess 
free energy functional $\mathcal{F}_\text{exc}$ 
by the non-local ``White Bear mark II'' functional for hard spheres 
\cite{Hansen_Goos_2006} (as in previous work, we additionally apply a correction by 
\textsc{Tarazona} \cite{tarazona_prl84_2000}). This latter functional allows to describe hard-sphere interactions 
between particles 
and between particles and the walls by employing fundamental measure theory \cite{roth_jpcm22_2010}. 
We refer to this approach by FMT. For explicit expressions and for details on the calculation 
of the functionals (via \textsc{Picard} iterations and solving the \textsc{Poisson} 
equation) we refer to previous work \cite{haertel_jpcm29_2017}. 
Adding a \textsc{Stern} layer for PB and mPB is discussed in \cref{sec:addingSternLayer}.
Note that electrostatic interactions beyond mean-field are still neglected in our \textsc{Coulomb} functional. 

To determine the reversible heat, we go back to the previous result in \cref{eq:QrevI'b} now. 
In a canonical system the reversible heat produced while charging our model system from 
surface charge density $0$ to $\sigma$ 
can also be expressed as (see \cref{sec:integratedheatprod}) 
\begin{align} \label{eq:QrevII}
\mathcal{Q}_\text{rev}^\text{II}&= A\int\limits_0^{\sigma}\int\limits_0^L\ \psi(z) 
\frac{\dd}{\dd\sigma'}\left(\sum\limits_{i}z_i \rho_i(z)\right)\,\dd z\,\dd\sigma' \,.
\end{align}
We introduce the superscript II to distinguish between our previous result in \cref{eq:QrevI'b} 
and this result in \cref{eq:QrevII}, 
where microscopic details enter explicitly through the density profiles of the system. 
Thus, this method is suitable to determine the reversible heat from DFT data, 
if calculations are performed for a canonical system. 

\subsection{MD simulations}\label{sec:mdSimulations}

As an additional approach, we study our system of interest through 
molecular dynamics (MD) simulations. 
For this purpose, we use the ESPResSo software package \cite{Weik2019} 
with the velocity \textsc{Verlet} algorithm for the propagation of the particles in our system. 
Hence, no real hard-sphere interaction can be used. Instead, we mimick the hard-core interactions 
by an extremely repulsive \textsc{Weeks-Chandler-Andersen} (WCA) potential 
\cite{weeks_jcp54_1971,*andersen_pra4_1971}, 
essentially a cut and shifted \textsc{Lennard-Jones} potential, that reads 
\begin{align}\label{eq:wca}
\frac{V_\text{WCA}(r)}{4\epsilon} &= \begin{cases}
\left(\frac{d_\text{LJ}}{r}\right)^{12} - \left(\frac{d_\text{LJ}}{r}\right)^{6}+\frac{1}{4} & \text{for } r \le 2^{1/6}d_\text{LJ}\,,\\
0 & \text{otherwise}\,. 
\end{cases}
\end{align}
In \cref{sec:simulationdetails} we explain our choice of the parameters $\epsilon$ 
and $d_\text{LJ}$ and verify that this choice yields neutral-sphere density profiles consistent with 
DFT (FMT) calculations. 
For the electrostatic interactions, ESPResSo provides the P3M method, a sophisticated \textsc{Ewald} method, 
as well as an electric layer correction (ELC) method to effectively remove the periodicity in one direction. 
We use both methods in a three-dimensional simulation box with periodic boundary conditions such that 
periodicity in the $x$ and $y$ directions account for the translational invariance of the system in those 
directions and the periodicity in $z$-direction is suppressed (see \cref{pic:setup}(b)). 

To model the effects of the charged walls, we first ensured that the EDLs were free 
of overlap such that we could run simulations with surface charges of opposite sign on both plates 
(see also discussion in \cref{sec:model}). Then we applied an additional constant electric field 
$E_z$ along the $z$-direction to all particles, which equals the field induced solely by the 
surface charges. 
The corresponding electrostatic potential difference between both walls 
for a given electric field strength $E_z$ is then obtained by 
\begin{align}
\Delta\Psi&=\frac{m_z}{\varepsilon_0\varepsilon_rA}-E_zL\,, 
\end{align}
where $m_z=e\int z(z_+\rho_+(z)+z_-\rho_-(z))\,\dd z$ is the electric dipole moment of the collective distribution of the ions along 
the $z$-direction.

\subsection{$\mathcal{Q}_\text{rev}$ depends sensitively on the boundary conditions of the charging process}

Next, we show that $\mathcal{Q}_\text{rev}/W_\text{el}$ differs dramatically between charging processes at either fixed $\mu_\pm$, $\bar{\rho}_\pm$, or $N_{\pm}$. 
	For illustrative purposes, we use the {\sc Gouy-Chapman} solution to the {\sc Poisson-Boltzmann} equations in this section: This solution allows for (semi) analytical expressions for $\mathcal{Q}_\text{rev}/W_\text{el}$ under the three above thermodynamic conditions.
	Charging processes at constant $\mu_\pm$ and $N_\pm$ are most easily treated in the  well-known grand canonical and canonical ensembles, respectively.
	We refer to the charging at fixed bulk densities  $\bar{\rho}_\pm$ as \emph{semi-canonical}, as $\mathcal{Q}_\text{rev}$ generated under this thermodynamic condition turns out to be close to the heat generated in large canonical systems.

\subsubsection{Recap of the {\sc Gouy-Chapman} solution}

\textsc{Gouy} and \textsc{Chapman} solved the {\sc Poisson-Boltzmann} equations 
for a setup of
one planar charged hard wall next to an infinite reservoir of $1:1$ electrolyte for which 
$\bar{\rho}_{+}=\bar{\rho}_{-}$. 
The solution reads \cite{gouy_1910,chapman_1913} 
\begin{subequations}\label{eq:gouy_chapman_densities}
	\begin{align}
		\phi(z) &= 4 \arctanh\left[\exp\left(-\frac{z}{\lambda_\text{D}}\right)\tanh\left(\frac{\Phi}{4}\right)\right]\,,\label{eq:gouy_chapman_densities_potential}\\
		\rho_\pm(z) &= \bar{\rho}_\pm\exp\left[\mp \phi(z)\right]\,,
	\end{align}
\end{subequations}
where $\lambda_\text{D}$ is the \textsc{Debye} length with 
$\lambda_\text{D}^{-2}=4\pi\lambda_\text{B}\sum_iz_i\bar{\rho}_i$ and 
$\lambda_\text{B}=e^2/(4\pi\varepsilon_0\varepsilon_r\kB T)$ is the \textsc{Bjerrum} length. 
Note that \cref{eq:gouy_chapman_densities} can be easily reformulated for a 
general $|z_i|:|z_i|$ electrolyte and so can the results that we derive below with \cref{eq:gouy_chapman_densities}. 

From \cref{eq:gouy_chapman_densities_potential} follows the surface charge density with {\sc Gauss}'s law as
\begin{align}
	\sigma &= \bar{\sigma} \sinh\left(\frac{\Phi}{2}\right)\,,\label{equ:GC_surface_charge_density}
\end{align}
where $\bar{\sigma}=4\ld \bar{\rho}_{+}$. 
Moreover, inserting \cref{eq:gouy_chapman_densities} into \cref{eq:OmegaDFT} 
we find
(cf.~Eqs. (24)~and (25) of Ref.~\cite{THEODOOR199061}) 
\begin{subequations}\label{eq:GCtotalOmega}
	\begin{align}
		\Omega^{\text{GC}} &= \Delta\Omega_\text{el}^{\text{GC}} + \Delta\Omega_\text{ent}^\text{GC} -pV,\label{eq:GCtotalOmegatot} 
		\intertext{with the bulk pressure $p=\kB T\sum_i\bar{\rho}_i$ of an ideal gas and}
		\frac{\Delta\Omega^{\text{GC}}_{\text{el}}}{A \kB T} &= \bar{\sigma} \left[\cosh\left(\frac{\Phi}{2}\right) - 1\right]\label{eq:GCtotalOmegaEL},\\
		\frac{\Delta\Omega^{\text{GC}}_{\text{ent}}}{A \kB T} &= \bar{\sigma}\left[3 - 3\cosh\left(\frac{\Phi}{2}\right)
		+ \Phi\sinh\left(\frac{\Phi}{2}\right)\right]\,.\label{eq:GCtotalOmegaent}
	\end{align}
\end{subequations}
Here, $\Omega^\text{GC}$ was partitioned into an entropic contribution $\Delta\Omega_{\rm ent}^\text{GC}$ 
and an energetic contribution $\Delta\Omega_{\rm el}^\text{GC}$ \cite{THEODOOR199061}.
The energetic contribution stems from the mean-field \textsc{Coulomb} functional $\mathcal{F}_{C}$, 
while the contribution $\Delta\Omega_{\rm ent}^\text{GC}-pV$ equals $\mathcal{F}_\text{id}-\sum_i \mu_i N_i$. 
However, as pointed out by \textsc{Overbeek}, $\mathcal{F}_{C}$ only yields purely energetic terms if a 
constant, in particular temperature\blue{-}independent, dielectric constant is used (see also \cref{appendix:QrevI}).

We will demonstrate in the next subsection that $\Delta\Omega_{\rm ent}^\text{GC}$ as defined in \cref{eq:GCtotalOmegaent} does not fulfill the 
correspondence given in \cref{eq:qref_omega_ent} (cf. \cref{eq:GCgc}). 
However, as it turns out, $\Delta\Omega_\text{ent}^\text{GC}$ is closely related to the entropic contribution to 
the free energy in canonical systems (cf. \cref{sec:254}).

\subsubsection{ {\sc Gouy-Chapman} at fixed $\mu_{\pm}$ (grand canonical)}

Inserting $\Omega^{\rm GC}$ from \cref{eq:GCtotalOmega} into \cref{eq:Qrev,equ:entropy_difference} to obtain 
$\mathcal{Q}_\text{rev}$, we find
\begin{align}\label{eq:GCgc}
	\frac{\mathcal{Q}_\text{rev}^\text{GC}}{A\kB T  \bar{\sigma}} &=\left(\frac{9}{2}-\frac{\mu_+}{\kB T}\right)\left[\cosh\left(\frac{\Phi}{2}\right)-1\right]
	- \Phi\sinh\left(\frac{\Phi}{2}\right).
\end{align}
In \cref{appendix:QrevI} we show that inserting $\Phi(\sigma)$ (as follows from inverting \cref{equ:GC_surface_charge_density}) into \cref{eq:QrevI'} yields the same expression for $\mathcal{Q}_\text{rev}^\text{GC}$.

\Cref{fig:heatensembles}(a) shows the ratio $\mathcal{Q}_\text{rev}^\text{GC}/W_\text{el}>0$ of reversible heat and electric work for grand canonical charging. 
	Here, we used the atomic mass $\unit[23]{u}$ corresponding to sodium to determine $\mathcal{Q}_\text{rev}^\text{GC}$ 
	and we used \cref{equ:work_pot_equivalence,eq:GCtotalOmegatot} to determine $W_\text{el}$ from $\Delta\Omega^{\text{GC}}$.
We observe $\mathcal{Q}_\text{rev}^\text{GC}/W_\text{el}>0$ up to $\Delta\Psi\approx\unit[0.9]{V}$, 
suggesting that heat flows into the system during charging, contradicting the experimental 
findings of Ref.~\cite{janssen_prl119_2017}. 
This positive ratio is caused by the net ion adsorption in both electrodes 
$\Delta N=4 A\bar{\sigma} \left[\cosh\left(\Phi/2\right) - 1\right]\geq 0$ 
within the system during charging when 
more counterions are attracted than coions are expelled; see also \cref{equ:particle_numbers} 
and \cref{appendix:canonical_GC}. 
Yet, an entropy contribution from increasing particle numbers is unlikely to have occured in the experiments of Ref.~\cite{janssen_prl119_2017}: 
	The system of porous electrodes and electrolyte reservoir used there, though certainly large, was closed and, hence, canonical.
	We conclude that the reversible heat $\mathcal{Q}_\text{rev}^\text{GC}$, with its uncommon explicit dependence on the ionic chemical potentials (hence on  \textsc{Planck}'s constant and ionic mass alike), is not relevant for the experimental setup of Ref.~\cite{janssen_prl119_2017}.

\subsubsection{ {\sc Gouy-Chapman} at fixed $N_{\pm}$ (canonical)}
Going from a grand canonical to a canonical description, the total numbers of particles per species $N_\pm^\text{tot}$ are kept fixed during charging rather than the  chemical potentials $\mu_\pm$. Hence, tracing the system states in a two-dimensional $(N_\pm^\text{tot},\mu_\pm)$ diagram during charging, 
grand canonical systems move along lines of constant $\mu_i$ whereas canonical systems move along 
lines of constant $N_\pm^\text{tot}$. As our model of the supercapacitor consists out of two charged hard walls for each electrode, the total numbers of both cations $N_{+}^\text{tot}$ and anions $N_{-}^\text{tot}$ read
\begin{align}
	N_{\pm}^\text{tot}=
	\left[ A\int\limits_0^L\rho_\pm(z)\,\dd z \right]_{
		\begin{matrix}\text{\tiny at pos.}\\\text{\tiny electrode}\end{matrix}
	}
	\hspace{-0.2cm}
	+ \left[ A\int\limits_0^L\rho_\pm(z)\,\dd z \right]_{
		\begin{matrix}\text{\tiny at neg.}\\\text{\tiny electrode}\end{matrix}
	} \hspace{-0.2cm} \,. \label{equ:particle_numbers}
\end{align}

We consider systems whose EDLs are free of overlap:
	the smallest $L=\unit[10]{nm}$ used in this subsection is still much larger than $\lambda_\text{D}\approx\unit[0.31]{nm}$.
Then, we determined $W_\text{el}$ during canonical charging with the \textsc{Gouy-Chapman} solution as follows.
For a given $\Phi$, we inserted $\rho_{\pm}(z)$ from \cref{eq:gouy_chapman_densities} into \cref{equ:particle_numbers} and varied $\bar{\rho}_{\pm}$ until the prescribed $N_\pm^\text{tot}$ was attained. 
	Clearly, the bulk densities $\bar{\rho}_\pm$ decrease while $\Phi$ increases at fixed $N_\pm^\text{tot}$  \cite{boon2011blue}.
	For each combination of $\Phi$ and  $\bar{\rho}_{\pm}$, we find $\sigma$ with \cref{equ:GC_surface_charge_density}, after which $W_\text{el}$ follows from \cref{eq:work} straightforwardly (see also \cref{appendix:canonical_GC}).
Next, to determine the heat $\mathcal{Q}_\text{rev}$ produced during canonical charging, 
	we are confronted with the problem that DFT is
formulated in the grand canonical ensemble. 
However, since our system is assumed to be 
infinite along the in-plane directions, the equivalence between the thermodynamic potentials 
(here $\Omega$ and $F$) holds true. 
We may thus perform a \textsc{Legendre} transform to obtain the free energy of our system as $F=\Omega+\sum_i\mu_iN_i$, where we use $\Omega^{\text{GC}}$ (\cref{eq:GCtotalOmega}) in place of $\Omega$.
Numerically calculating the derivative of the free energy with respect to temperature then yields the reversible heat. 

\Cref{fig:heatensembles} shows ratios $-\mathcal{Q}_\text{rev}/W_\text{el}$ of reversible heat and electric work for canonical systems of different lengths $L$. 
Here, the expected sign, corresponding to heat flowing out 
of the system during charging, is obtained. 
We also note that, though all systems considered are free of EDL overlap, $-\mathcal{Q}_\text{rev}/W_\text{el}$ depends markedly on $L$.
	This is because, the smaller the system, the faster $\bar{\rho}_{\pm}$ decreases during canonical charging.
The connected 
reservoir in the experiment of 
	Ref.~\cite{janssen_prl119_2017} being large in comparison to the volume filled by EDLs and desalination of the bulk 
being negligible during charging brings up the question as to how the ratio $-\mathcal{Q}_\text{rev}/W_\text{el}$ behaves
in the limit of large systems where $L/\lambda_\text{D}\to\infty$. 
Based on the arguments of the previous subsection, we do not expect $-\mathcal{Q}_\text{rev}/W_\text{el}$ to be the same in grand canonical and canonical processes in this limit.

\subsubsection{ {\sc Gouy-Chapman} at fixed $\bar{\rho}_\pm$ (semi-canonical)}\label{sec:254}
Reference~\cite{janssen_prl119_2017} found an expression for the $\mathcal{Q}_\text{rev}$ using \cref{eq:QrevII}, which holds for canonical systems only, inserting, however, {\sc Gouy-Chapman} density profiles pertaining to a grand canonical system.
Interestingly, their expression for $\mathcal{Q}_\text{rev}^\text{II}$ also follows from combination of \cref{eq:qref_omega_ent,eq:GCtotalOmegaent}.
Now, an identical expression for $\mathcal{Q}_\text{rev}^\text{I}$ can be obtained from \cref{eq:Qrev1,eq:defQrev} \emph{if} the partial derivatives therein  
are carried out not at fixed $\mu_i$ (imperative in grand canonical settings), but 
for constant $\bar{\rho}_\pm$, that is, a $\bar{\rho}_\pm$ independent of $T$, 
$\sigma$, and hence $\Phi$.

Importantly, at fixed $\bar{\rho}_{\pm}$, the chemical potentials $\mu_i$ vary with $T$ and that the particle numbers $N_\pm$ vary with $\Phi$.
Hence, fixed-$\bar{\rho}_\pm$ charging is neither grand canonical nor canonical, and we call it 
``semi-canonical'' instead.
Meanwhile, as the ionic density profiles for given $\bar{\rho}_\pm$ are the same in grand canonical and semi-canonical systems, they have the same $\Phi(\sigma)$-relation and, through  \cref{eq:work}, the same $W_\text{el}$ as well.

\begin{figure}
	\centering
	\includegraphics[scale=1.0]{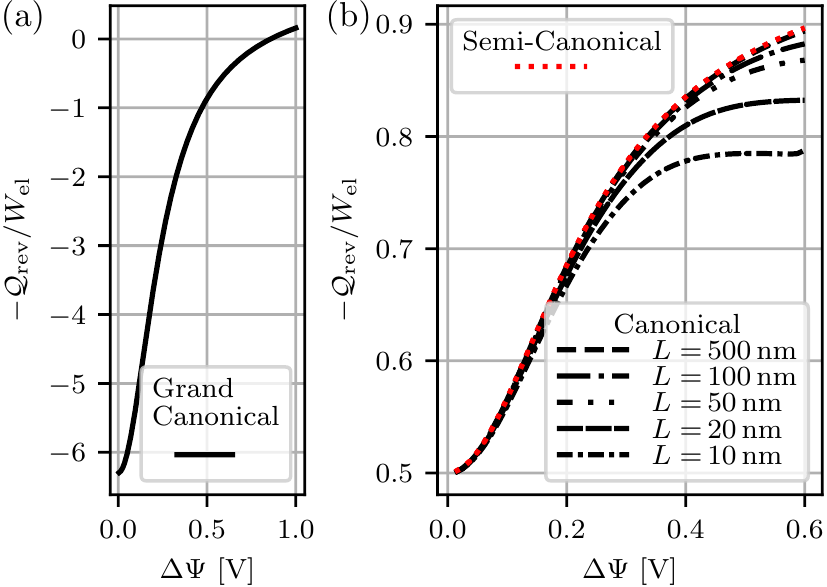}
	\caption{ 
		Ratio $-\mathcal{Q}_\text{rev}^\text{GC}/W_\text{el}$ of negative reversible heat and 
		electric work during isothermal charging as a function of applied potential $\Delta\Psi$ 
		calculated with the \textsc{Gouy-Chapman} solution 
		(a) in a grand canonical setting at fixed $\mu_{\pm}$ and 
		(b) in a canonical setting at fixed $N_{\pm}$ for several porewidths $L$. 
		Panel (b) also shows the ratio $\Delta\Omega^{\text{GC}}_{\text{ent}}/\Delta\Omega^{\text{GC}}$ 
		in a semi-canonical setting as follows from \cref{eq:GCtotalOmega}, 
		which corresponds to a fixed $\bar{\rho}_{\pm}$. }
	\label{fig:heatensembles}
\end{figure}

In \cref{fig:heatensembles} we plot the ratio of reversible heat and electric work 
obtained from the \textsc{Gouy-Chapman} solution via \cref{eq:GCtotalOmega}. As discussed above, 
the ratio $\Delta\Omega_{\rm ent}^\text{GC}/\Delta\Omega^\text{GC}$ in this semi-canonical system 
indeed represents the limiting ratio in the canonical system for 
increasing amounts of connected bulk and, thus, increasing $L$.
It is astonishing that calculations carried out for this semi-canonical process at fixed $\bar{\rho}_\pm$ 
do not only reflect the conditions of the experiment much better but also simplify 
calculations (e.g. \cref{eq:Qrevexplicit,eq:QrevSimple}). 

\subsubsection{Conclusion}

As demonstrated using the \textsc{Gouy-Chapman} solution, the ratio $\mathcal{Q}_\text{rev}/W_\text{el}$ 
depends strongly on the used thermodynamic conditions. 
Next to conventional grand canonical and canonical charging processes, we introduced a third process, namely a \emph{semi-canonical} charging process at constant $\bar{\rho}_\pm$. 
This process mimics a charging process in a system connected to an infinite bulk, such that the system in 
combination with the bulk is canonical.
Even though the density profiles for different $\sigma$ or $\Phi$ but constant $T$ 
are the same as in a grand canonical system,
the reversible heat produced during semi-canonical charging resembles the heat generated in a canonical system instead.
Importantly, this semi-canonical process reflects the conditions of 
the experiment as described in Ref.~\cite{janssen_prl119_2017} best and, accordingly, it is used 
in the following. 

\section{Results}

We consider a parameter set corresponding to the experiment of Ref.~\cite{janssen_prl119_2017}: 
$T=\unit[300]{K}$, $\bar{\rho}_\pm=\unit[1]{M}$, $d=\unit[0.68]{nm}$, and 
$\varepsilon_r=80$ (in all approaches, including SPM) resulting in a \textsc{Bjerrum} length of 
around $\lambda_\text{B}=\unit[0.696]{nm}$. 
Note that $d$ only enters in $\mathcal{F}_\text{hs}$ and $\mathcal{F}_\text{lg}$. 
Moreover, we now set $L=\unit[10]{nm}$ and discuss narrower pores later in \cref{sec:newsettings}. 

\subsection{Check of $\mathcal{Q}_\text{rev} = -\Delta\Omega_\text{ent}$ (\cref{eq:qref_omega_ent}) }

\begin{figure}
  \centering
  \includegraphics[scale=1]{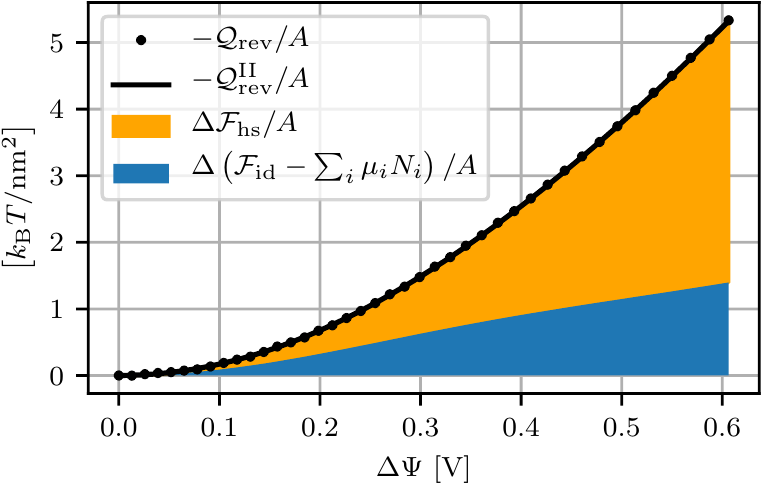}
  \caption{Negative reversible heat $-\mathcal{Q}_\text{rev}$ per area $A$ obtained via 
    \cref{eq:Qrev} (dots) and \cref{eq:QrevII} (solid line) from FMT RPM during an isothermal 
    charging process. 
    In addition, we show the change of the contributions 
    $\mathcal{F}_\text{hs}$ (upper orange area) and $\mathcal{F}_\text{id}-\sum_i\mu_i N_i$ (lower blue area) 
    to the grand potential. }
	\label{fig:heatContributions}
\end{figure}

In \cref{fig:heatContributions}, we show results from FMT RPM for the reversible 
heat $\mathcal{Q}_\text{rev}^\text{I}$ and $\mathcal{Q}_\text{rev}^\text{II}$. Both calculations via \cref{eq:Qrev} (dots) and 
\cref{eq:QrevII} (solid line) clearly yield the same result numerically. 
We further check numerically whether the free-energy contribution of the hard-sphere interaction 
within FMT adds to the entropic contribution to the grand 
potential. For this purpose, we calculate the contributions $\mathcal{F}_\text{hs}$ and 
$\mathcal{F}_\text{id}$ with the density profiles from FMT; 
we calculate the particle numbers via \cref{equ:particle_numbers}. 
The two resulting terms $\Delta\mathcal{F}_\text{hs}$ and 
$\Delta(\mathcal{F}_\text{id}-\sum_i\mu_i N_i)$, 
shown with colored areas in \cref{fig:heatContributions}, add up to precisely %$-\mathcal{Q}_\text{rev}$. 
$\Delta\Omega_\text{ent} = -\mathcal{Q}_\text{rev}$ (\cref{eq:qref_omega_ent}). 
Thus, as expected, the contribution of the hard-sphere excess term goes completely 
into the entropic part. 
We performed the same checks in the mPB model. Again grouping the volume exclusion 
term ($\mathcal{F}_\text{lg}$) into the $\Delta\Omega_\text{ent}$ term, we verified 
$\Delta\Omega_\text{ent} = -\mathcal{Q}_\text{rev}$ also for mPB. 

In conclusion, in two new cases we have numerically verified that 
$\Delta\Omega_\text{ent} = -\mathcal{Q}_\text{rev}$ only holds if the excess functional accounting for steric 
interactions is grouped into the entropic contribution $\Delta\Omega_\text{ent}$ of the grand potential. 
From hereon, we prefer to speak about $\mathcal{Q}_\text{rev}/W_\text{el}$ instead of 
$\Delta\Omega_{\text{ent}}/\Delta\Omega$, although the latter has been used in previous 
work \cite{janssen_prl119_2017,biesheuvel2007counterion}. 
This is because both $\mathcal{Q}_\text{rev}$ and $W_\text{el}$ are unambiguously defined 
in \cref{eq:work,eq:Qrev,eq:QrevII} and can be measured experimentally. 

\subsection{$\mathcal{Q}_\text{rev}/W_\text{el}$ within theoretical approaches}

\begin{figure*}
	\centering
	\includegraphics[scale=1.0]{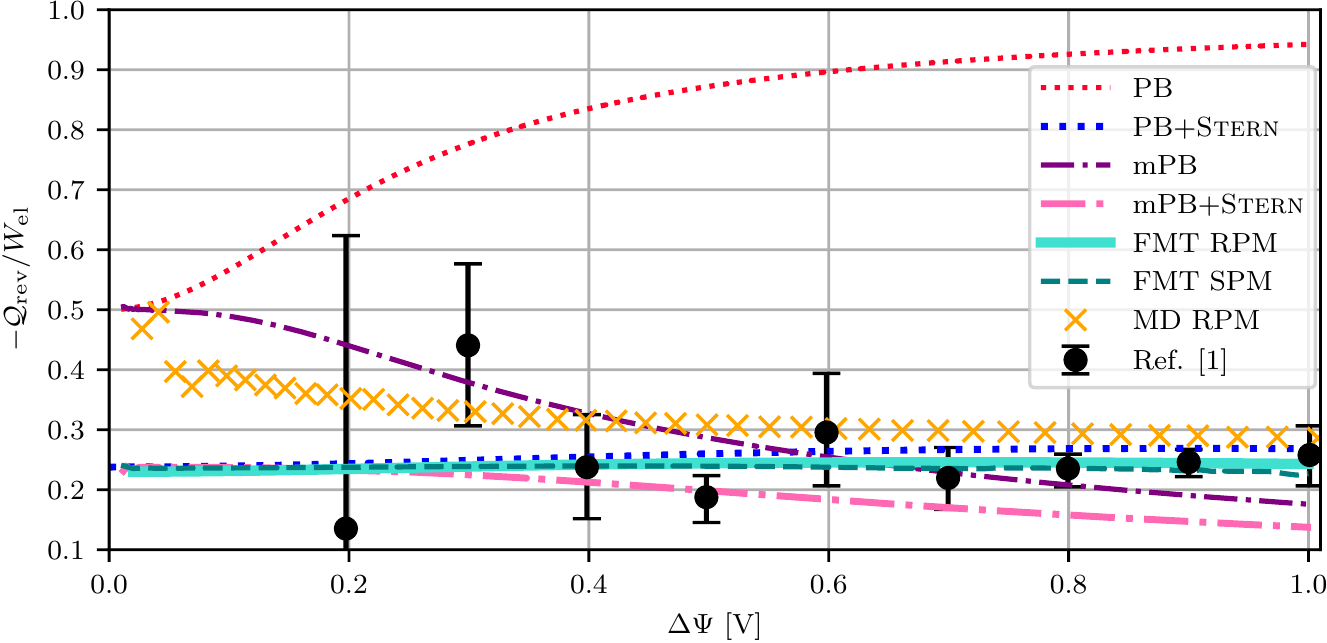}
	\caption{
    Ratio $-\mathcal{Q}_\text{rev}/W_\text{el}$ of negative reversible heat and electric work during isothermal 
    charging as a function of applied potential $\Delta \Psi$ 
    for PB without (red dotted line) and with \textsc{Stern} layer (thick blue dotted line), 
    mPB without (purple dash-dotted line) and with \textsc{Stern} layer (thick pink dash-dotted line), 
    FMT RPM (cyan solid line), 
    FMT SPM (teal dashed line), and MD simulations (orange crosses), all determined for 
    the semi-canonical process at fixed $\bar{\rho}_\pm=1$ M with 
    $L=10$ nm (EDLs are free of overlap for $L\geq 4$ nm), $d=0.68$ nm, and $|z_\pm|=1$. 
    Also shown with black dots are the experimental results from Ref.~\cite{janssen_prl119_2017}. }
	\label{fig:heatWorkRatios}
\end{figure*}

In \cref{fig:heatWorkRatios}, we show the ratio $-\mathcal{Q}_\text{rev}/W_\text{el}$ obtained 
from the different methods introduced before as well as the experimental data of Ref.~\cite{janssen_prl119_2017}. Explicitly, we 
compare the ratios obtained via PB without (red dotted line, same data as in \cref{fig:heatensembles}(b)) 
and with \textsc{Stern} layer (thick blue dotted line), mPB without (purple dash-dotted line) 
and with \textsc{Stern} 
layer (thick pink dash-dotted line), FMT RPM (cyan solid line), FMT SPM (teal dashed line), and 
MD simulations (orange crosses).

First, we notice that PB, the only description where all steric interactions among the particles 
are neglected, deviates from all other curves in the way that 
$\mathcal{Q}_\text{rev}\to -W_\text{el}$ at large potentials. At small applied potentials, we have
$\mathcal{Q}_\text{rev}=-W_\text{el}/2$. This is all in perfect agreement with earlier 
descriptions by \textsc{Overbeek} \cite{THEODOOR199061}. 
Second, as speculated in Ref.~\cite{janssen_prl119_2017}, mPB theory describes the experimental 
data much better than PB theory. 
PB and mPB coincide at small applied potentials, which is understood from their equal leading-order 
expansion in $\Delta\Psi$, the \textsc{Debye-H\"uckel} equation \cite{kralj1996simple}. 
Conversely, for $\Delta\Psi$ much beyond the thermal voltage,
$-\mathcal{Q}_\text{rev}/W_\text{el}$ radically changes: 
PB predicts the ratio to rise to $1$ for large potentials, 
while mPB predicts this ratio to decrease with increasing potential instead. 
A similar qualitative change upon accounting for steric repulsions was found in 
Ref.~\cite{biesheuvel2007counterion}. 
Similar conclusions for PB and mPB also hold when we add a \textsc{Stern} layer, as explained in 
\cref{sec:addingSternLayer}. Accounting for a \textsc{Stern} layer, however, dramatically 
alters the ratio $-\mathcal{Q}_\text{rev}/W_\text{el}$ at low applied potentials. Interestingly, the 
value of $-\mathcal{Q}_\text{rev}/W_\text{el}$ around $0.24$ agrees well with the more sophisticated 
FMT approaches that we discuss now. 

We see that the predictions of FMT both for RPM and SPM are almost equal and agree 
with all experimental data within two standard deviations. 
The similarity of the RPM and SPM results, however, does not mean that solvent 
properties do not affect $-\mathcal{Q}_\text{rev}/W_\text{el}$. 
For example, the SPM does not account for dipolar interactions within water, 
which might influence $-\mathcal{Q}_\text{rev}/W_\text{el}$.
At large potentials, RPM and SPM predictions for the ratio $-\mathcal{Q}_\text{rev}/W_\text{el}$ 
are similar to those from mPB, but RPM and SPM predict this ratio to be roughly constant, 
while mPB predicts this ratio to still decrease with increasing potential. 
Interestingly, PB with \textsc{Stern} layer predictions are also similar to those from FMT, even for large 
potentials. At small 
potentials, RPM and SPM deviate significantly from $\mathcal{Q}_\text{rev}=-W_\text{el}/2$, 
as predicted by the approaches without \textsc{Stern} layer (PB and mPB).  
In both RPM and SPM particles cannot get closer to the wall than half a particle diameter, 
introducing a \textsc{Stern}-like layer, whereas in PB and mPB particles can get arbitrarily close to the wall. 

Further, we performed MD simulations for $\bar{\rho}_\pm=1$ M with $600$ particles per species in a box of 
$\left(\unit[10]{nm}\right)^3$. 
The MD simulations give access to the (equilibrium) internal energy $U$ and the electric work 
$W_\text{el}$ done to the system (see \cref{eq:work}), 
from which the heat flowing into the system follows as 
$\mathcal{Q} = \Delta U-W_\text{el}$\,. 
In \cref{fig:heatWorkRatios}, we see that MD (orange crosses) predicts slightly larger 
values for $-\mathcal{Q}_\text{rev}/W_\text{el}$ than FMT, and mostly describes the experiment worse. 
We performed a convergence analysis for the parameter $L$ by regarding this ratio for different system lengths. 
We found that the statistical error for potential differences $\Delta\Psi\ge\unit[0.2]{V}$ 
is smaller than the used marker size and hence negligible compared to the experimental uncertainty. 

The deviation between MD and FMT predictions can have different reasons: most probably, the 
approximate mean-field functional used for the electrostatic interactions in the FMT approach 
simply does not capture crucial contributions. For instance, it is known that inaccurate approaches 
like the FMT approach predicts qualitatively wrong adsorption to weakly charged walls in the 
RPM \cite{gillespie_jpcm17_2005}. Another, 
similar reason, could be the treatment of image charges in both MD and FMT. While image charges 
are not captured in our MD, it is not yet understood whether they are captured in the ensemble averaged 
DFT approach. 
Nevertheless, from the agreement between FMT and PB with \textsc{Stern} layer we conclude that the 
complex structure of the ion density profiles near the surface of the charged hard wall, as predicted only by 
the FMT approaches and MD, is less important than the steric interaction between the charged wall 
and the ions represented by the \textsc{Stern} layer.

\subsection{Influence of pore size, ionic diameter, bulk density, and valencies on $\mathcal{Q}_\text{rev}/W_\text{el}$}\label{sec:newsettings}

Using the FMT RPM approach, we discuss how different parameters affect $-\mathcal{Q}_\text{rev}/W_\text{el}$. 

\begin{figure}[h]
	\centering
	\includegraphics[scale=1]{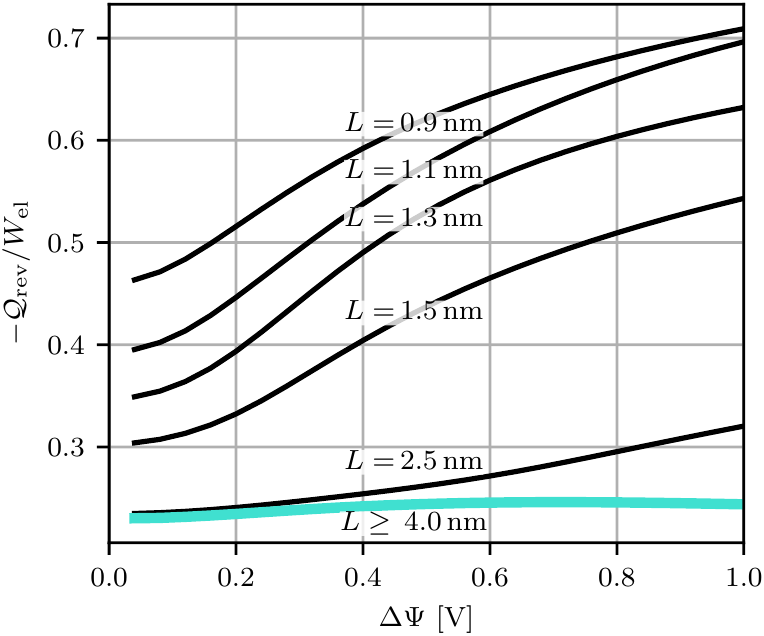}
	\caption{Ratio $-\mathcal{Q}_\text{rev}/W_\text{el}$ for FMT RPM as in 
		\cref{fig:heatWorkRatios}, but for different system lengths $L$. }
	\label{fig:OmegaratioL}
\end{figure}

\subsubsection{Pore size $L$}

\Cref{fig:OmegaratioL} shows $-\mathcal{Q}_\text{rev}/W_\text{el}$ for several $L$ and all 
other parameters as in \cref{fig:heatWorkRatios}, for which, in particular, \mbox{$\ld\approx\unit[0.31]{nm}$}. 
For interacting EDLs obtained for small pore sizes around $L\sim \ld$, one can see a 
rapid increase of $-\mathcal{Q}_\text{rev}/W_\text{el}$ with decreasing $L$. 
This finding is relevant to many supercapacitor experiments, where pores in electrodes 
are nanometer sized and, hence, strong EDL overlap can be expected. 
For systems larger than $L=\unit[4]{nm}$ no further effects from $L$ on 
$-\mathcal{Q}_\text{rev}/W_\text{el}$ can be seen because the EDLs decay almost completely 
within half a system length.
As stated earlier, to get the same ratio of pore volume to surface area as in the experiment, 
a pore size of $L=\unit[1.6]{nm}$ must be used. For such a pore size the agreement between the 
experiment and the FMT RPM curve would be much worse. However, the effective diameter of the ions 
used here includes some contributions due to hydration shells. These hydration shells might be 
partly shed when ions get adsorbed at the electrode causing some effective increase of pore size 
due to shrinking effective ion sizes. This issue cannot be completely resolved here with the 
excess functional $\mathcal{F}_\text{exc}$ that we use and needs further investigation. 

\subsubsection{Ionic diameter $d$}

\begin{figure}[h]
  \centering
  \includegraphics[scale=1]{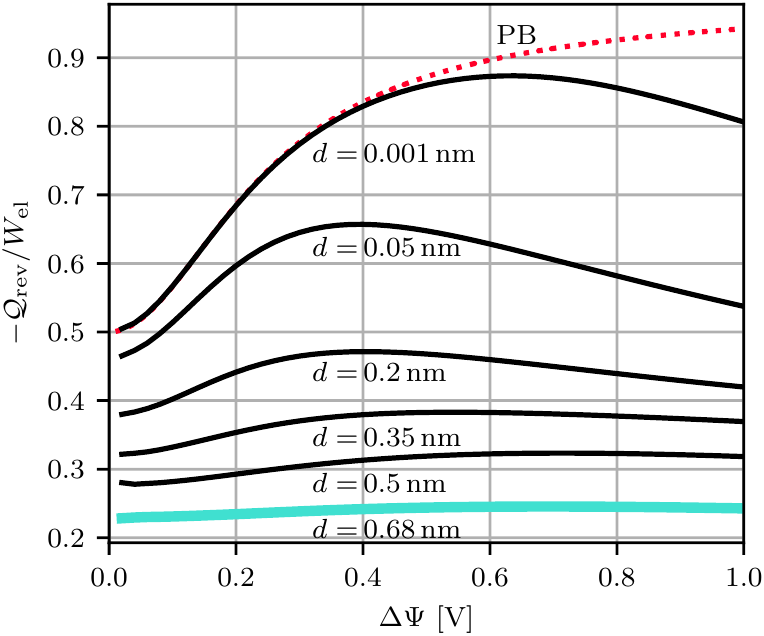}
  \caption{Ratio $-\mathcal{Q}_\text{rev}/W_\text{el}$ for FMT RPM as in 
    \cref{fig:heatWorkRatios}, but for different particle diameters $d$. 
    For small $d$ this ratio approaches PB predictions (red dotted line). }
	\label{fig:Omegaratiosigma}
\end{figure}

\Cref{fig:Omegaratiosigma} shows $-\mathcal{Q}_\text{rev}/W_\text{el}$ for several $d$ and 
all other parameters as in \cref{fig:heatWorkRatios}. We also show the analytical 
\textsc{Gouy-Chapman} solution to the \textsc{Poisson-Boltzmann} equations for point charges that follows 
from \cref{eq:GCtotalOmega}. We see that, 
with $d\to0$, the RPM results move progressively towards the PB predictions. 
Notably, however, even for (unphysically) small $d$, the RPM qualitatively differs 
from PB as it does not approach 1 but rather decreases at large $\Delta\Psi$.
We interpret these results as follows:
The change in entropy upon charging is 
associated with the increasing order in the system when ions separate. 
As PB predicts larger $-\mathcal{Q}_\text{rev}/W_\text{el}$ than RPM in 
\cref{fig:heatWorkRatios}, the change in entropy upon charging is strongest for point charges. 
Steric interactions of the hard spheres in the RPM model counteract this trend 
to order and thus decrease the ratio $-\mathcal{Q}_\text{rev}/W_\text{el}$. 

\subsubsection{Bulk density $\bar{\rho}_\pm$}

\begin{figure}[]
  \centering
  \includegraphics[scale=1]{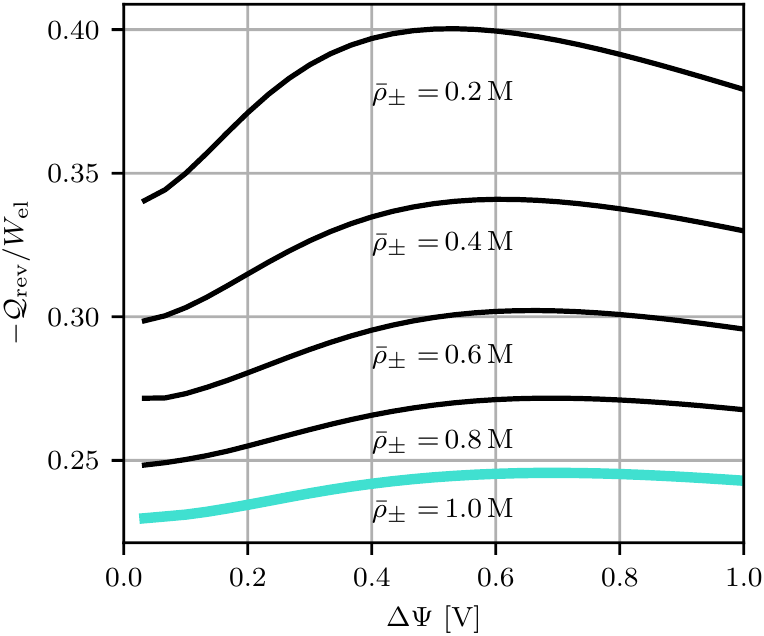}
  \caption{Ratio $-\mathcal{Q}_\text{rev}/W_\text{el}$ for FMT RPM as in 
    \cref{fig:heatWorkRatios}, but for different bulk concentrations $\bar{\rho}_\pm$. }
	\label{fig:OmegaratioC}
\end{figure}

\Cref{fig:OmegaratioC} shows $-\mathcal{Q}_\text{rev}/W_\text{el}$ for several $\bar{\rho}_\pm$ 
and all other parameters as in \cref{fig:heatWorkRatios}. 
Clearly, $-\mathcal{Q}_\text{rev}/W_\text{el}$ decreases with increasing $\bar{\rho}_\pm$. We have verified 
that $|\mathcal{Q}_\text{rev}|$ decreases with increasing $\bar{\rho}_\pm$ as well, 
as was found in experiments \cite{zhang_ta636_2016}. 
Note that in the RPM a phase transition would be expected at a packing fraction roughly above $0.45$ 
that corresponds to a bulk concentration of $\bar{\rho}_\pm=5.4666$ nm$^{-3}=9.1$ M 
\cite{hynninen_prl96_2006}, which is far from our system at $1$ M. 
Further, the reduced temperature $T^*=d/\lambda_\text{B}=0.977$ in our system is much larger 
than the temperatures where gas-liquid phase separation occur \cite{hynninen_prl96_2006}. 

\begin{figure}[]
	\centering
	\includegraphics[scale=1]{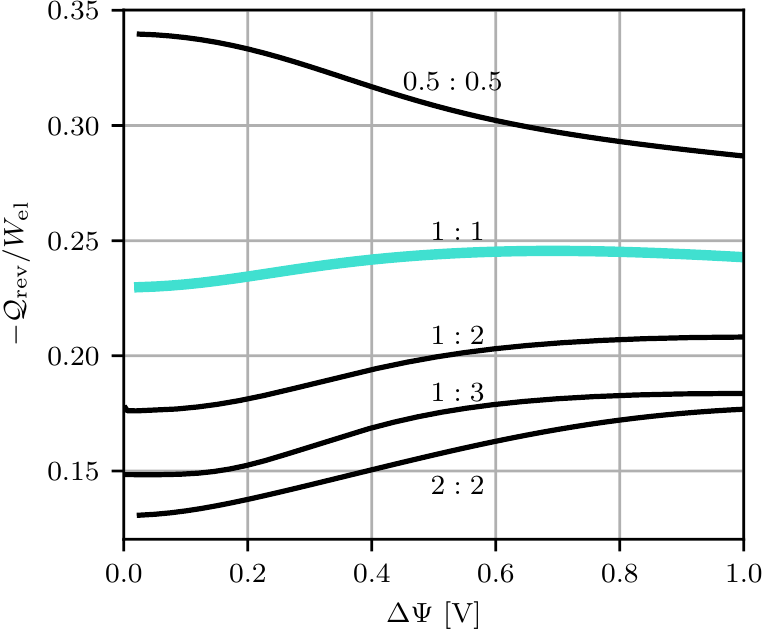}
	\caption{Ratio $-\mathcal{Q}_\text{rev}/W_\text{el}$ for FMT RPM as in 
		\cref{fig:heatWorkRatios}, but for different valencies $|z_+|:|z_-|$. }
	\label{fig:Omegaratiozi}
\end{figure}

\subsubsection{Valencies $z_{i}$}

\Cref{fig:Omegaratiozi} shows $-\mathcal{Q}_\text{rev}/W_\text{el}$ for several binary 
multivalent electrolytes. In order to get an electrically neutral bulk, the bulk 
densities $\bar{\rho}_\pm$ must be changed accordingly. 
We choose bulk densities such that $\bar{\rho}_{+}+\bar{\rho}_{-}=\unit[2]{M}$ to get the same 
total bulk particle density as before. All other parameters are as in \cref{fig:heatWorkRatios}. 
We see that $-\mathcal{Q}_\text{rev}/W_\text{el}$ decreases with the amount of total 
charge defined as $e\sum_i |z_i|\bar{\rho}_i$. 
Next to integer charges, we also studied one case of fractional charges, relevant 
for effective charges of larger molecules as they occur, for instance, in the description of 
ionic liquids \cite{roy_jpcb114_2010}. 

Note that, due to the simple form of the \textsc{Coulomb} interaction, a multiplicative factor 
for the valencies can be theoretically mapped onto different $\varepsilon_r$ or $T$. For example, 
the data with valencies $z_i=0.5$ corresponds to one with $z_i=1$ at a four times higher 
temperature. Note that also the surface charges, and hence the electric potential, would have to be rescaled.

\section{Discussion and Conclusions}

In this study, we calculated the ratio $-\mathcal{Q}_\text{rev}/W_\text{el}$ between the negative of 
the reversible heat emerging during isothermal charging processes and applied electric work for the 
RPM, SPM, and PB-type models of electrolytes. 
Our findings agree with the experimental results from Ref.~\cite{janssen_prl119_2017} 
and demonstrate the importance of ionic steric interactions with the 
charged wall to explain the experiments theoretically. 

First, we found that the ratio of reversible heat and electric work depends sensitively on the thermodynamic conditions. A semi-canonical process explains the experiments from Ref.~\cite{janssen_prl119_2017} best. 
This process describes charging in a system connected to a very large bulk reservoir such that 
the reservoir has constant bulk density during particle exchange, as in a grand canonical system. 
Simultaneously, the bulk density does not change with temperature such that the chemical 
potential does change, as in a canonical system. We demonstrated and discussed this finding 
for different ensembles and processes; for simplicity we used the \textsc{Gouy-Chapman} solution 
to PB theory. 

For the calculation of the reversible heat and electric work in the RPM, we used different 
theoretical approaches, namely classical DFT and MD simulations. To describe the ionic volume 
in DFT, we used a modified PB approach as well as a sophisticated approach using FMT. 
Furthermore, we performed calculations for point-like ions within 
\textsc{Poisson-Boltzmann} theory with and without a \textsc{Stern} layer to emphasize the importance of 
steric wall-ion interactions for a description of the experiments. While mPB and FMT predictions 
for $\mathcal{Q}_\text{rev}/W_\text{el}$ are similar for large $\Delta\Psi$, they deviate significantly 
around $\Delta\Psi=0$ V. Such differences would in principle be experimentally testable though 
the data of Ref.~\cite{janssen_prl119_2017} for $\Delta\Psi<0.2$ V have large uncertainty. 
Conversely, addition of a \textsc{Stern} layer to PB and mPB yields predictions 
in excellent agreement with FMT even at low potentials.
We point out that deviations between MD and FMT most probably arise from the inaccurate 
treatment of the electrostatic interactions in our used functional, but other sources like the
different treatment of image charges are possible as well.
Meanwhile, discrepancies between our theoretical calculations and the experimental data 
could have different causes: In our model we neglected adsorption and faradaic reactions 
and, further, the geometry in our model is an oversimplification of real porous electrodes. 
Moreover, the treatment of the solvent as a constant dielectric background means that we 
cannot describe the shedding of an ion's hydration shell when it enters an ultranarrow pore. 
Notwithstanding these reservations, our results point towards the important role of 
the finite size of particles to heat production experiments of capacitive systems. 

For this reason, we have further tested the importance of volume effects of the electrolyte 
solvent that is not contained in the RPM: We performed additional FMT 
calculations in the SPM, where steric interactions of solvent particles are treated 
explicitly; solvent particles are described as neutral hard spheres, while we retained the dielectric background of the RPM. We could 
not find significant deviations between our calculations for RPM and SPM. 

Anticipating more experimental data and having at hand a predictive theoretical approach, 
we further studied how the ratio $\mathcal{Q}_\text{rev}/W_\text{el}$ changes with pore 
size $L$, ion size $d$, salt concentration $\bar{\rho}_\pm$, and ionic valencies $z_\pm$. 
We found that the ratio depends sensitively on all these parameters.
Experimental heat production measurements should be able to pick these trends up. 

In the future, experiments for the ratio $\mathcal{Q}_\text{rev}/W_\text{el}$ could become 
a valuable tool to test aspects of EDL theories. However, the large experimental uncertainty 
of the available data at small applied potentials \cite{janssen_prl119_2017} hinders this at 
the moment. We thus hope that our study inspires more experimental work on this topic. 
Furthermore, our findings are of interest for applications, where EDLs change cyclically or 
where they are exploited in combination with temperature changes, for instance in heat-conversion 
processes.

Finally, DFT has the advantage to provide thermodynamic potentials innately. 
While DFT is limited to equilibrated systems, dynamical DFT allows to 
study non-equilibrium processes, for instance, fast (dis)charging of 
supercapacitors where \textsc{Joule} heating comes into play as well \cite{schmidt_jcp138_2013,schmidt_pre84_2011,*anero_jcp139_2013,lee_prl115_2015}. 
Such a framework could be of interest for several applications. 
For instance, recently the heat produced in nervous conduction has been found to contain a 
large fraction of reversible heat, while irreversible contributions are small, if existing 
at all \cite{deLichtervelde_pre101_2020}. Now, our work could guide studies, where the theoretical 
description of nervous conduction goes beyond ideal solutions and homogeneous bulk concentrations, 
as typically applied, and, thus, could shed new light on this fundamental neuroscience process.

\section*{Acknowledgements}

FG and AH acknowledge support by the state of 
Baden-Württemberg through bwHPC and the German Research Foundation (DFG) through grant no 
INST 39/963-1 FUGG (bwForCluster NEMO). 
The research leading to these results has received funding from the European Union's 
Horizon 2020 research and innovation programme under the Marie Sk\l{}odowska-Curie grant agreement No~801133. 
Finally, FG acknowledges funding by the 
Deutsche Forschungsgemeinschaft (DFG, German Research Foundation) - project number 430195928 - 
and AH acknowledges funding by the DFG - project number 406121234. 

\section*{Data availability statement}

The data that support the findings of this study are available from the corresponding author upon reasonable request. 

%\bibliography{literature/literature}

\begin{appendix}

\section{Simulation details}\label{sec:simulationdetails}

For the MD simulations, we had to set the WCA parameters entering \cref{eq:wca}. 
For the strength of the repulsion we used $\epsilon=100\kB T$, as in previous work \cite{haertel_jpcm27_2015}. 
Note that in literature many different methods to obtain an effective radius can be 
found \cite{andersen_pra4_1971,doi:10.1063/1.1701689}.
We set $d_\text{LJ}$ such that density profiles from DFT and MD agree well for a system of neutral hard spheres, which happens when 
$2^{1/6}d_\text{LJ}$ is about $1.4\%$ larger than the hard-sphere diameter.
To obtain this value, we demand that the first \textsc{Mayer} $f$-bond contribution for a 
hard-sphere system equals the one for the WCA potential \cite{hansen2013theory}. 

As the counterion density near the walls shoot up with the applied potential, 
one needs to check if the simulation box is sufficiently large to capture the spatial correlations along 
the lateral directions. Accordingly, we calculated the radial distribution function 
projected on the lateral plane for particles close to the wall. We
checked that the projected radial distribution function decays to zero within half a lateral 
box length. Further, we checked that the length of the system along the normal direction is 
long enough such that, as for DFT, the different EDLs do not interact nor desalinate 
the bulk. We found that a box of $\left(\unit[10]{nm}\right)^3$ with $600$ particles 
per species meets the desired conditions.
The particle numbers in the simulations are chosen such that
we find the bulk densities from DFT calculations in the center of the simulation box. This allows us to use oppositely instead of equally charged walls in our simulations. Hence, an homogeneous electric field can be used to mimic the effects of surface charges.

\section{Adding a \textsc{Stern} layer to the PB and mPB approach}\label{sec:addingSternLayer}

The finite size of ions affects both the ion-ion interaction as well as the ion-wall interaction. 
A simple way to account for finite ion size (in the ion-wall interaction) is through a \textsc{Stern} 
layer, which is a charge-free region reaching from the electrode surface into the electrolyte over the 
ionic radius $d/2$. From \cref{eq:poisson} and \textsc{Gauss}'s law 
$\partial_z \phi = -4\pi \lambda_\text{B} \sigma$ follows the potential difference 
over the \textsc{Stern} layer as $\Phi^{\rm \textsc{Stern}}=2\pi \lambda_\text{B} \sigma d$. 
Note that the same potential drop $\Phi^{\rm \textsc{Stern}}$ applies to PB and mPB theory.

In \cref{eq:GCtotalOmegaEL,eq:GCtotalOmegaent}, we expressed $\Omega_\text{el}^\text{GC}$ as a
function of the potential $\Phi$. 
Using, instead of \cref{equ:GC_surface_charge_density}, that 
\begin{align}\label{eq:potGCStern}
\Phi &= 2\arcsinh\left(\frac{\sigma}{\bar{\sigma}}\right) +2\pi\lambda_\text{B}\sigma d \,,
\end{align}
we can express $\Omega_\text{el}^\text{GC}$ as a function of $\sigma$ as
\begin{align}
   \Omega_\text{el}^\text{GC+\textsc{Stern}}(\sigma) &=
\Omega_\text{el}^\text{GC}(\sigma)
   + \frac{e^2\sigma^2Ad}{4\varepsilon_0\varepsilon_r} .
\label{eq:omegaStern}
\end{align}
As $\Omega^\text{GC}_\text{ent}(\sigma)$ is unaffected by the \textsc{Stern} layer 
we obtain the ratio 
$-\mathcal{Q}_\text{rev}/W_\text{el}$ as a function of $\sigma$ as 
\begin{align}
   -\frac{\mathcal{Q}_\text{rev}}{W_\text{el}}(\sigma)
   &=
\frac{\Omega_\text{ent}^{\text{GC}}(\sigma)}{\Omega_\text{ent}^{\text{GC}}(\sigma)+\Omega_\text{el}^\text{GC+\textsc{Stern}}(\sigma)}
.
\end{align}
Using this result together with \cref{eq:potGCStern} to obtain the potential $\Phi$, we obtain the 
thick dotted blue line shown in \cref{fig:heatWorkRatios}. 

\section{Reversible heat from integrated heat production}\label{sec:integratedheatprod}

For a thermodynamical (dis)charging process, the difference in heat $\delta\mathcal{Q}$ can 
be calculated from the difference 
in internal energy $\dd U$ and the work $\delta W$ done during the process as 
\begin{align}
	\delta \mathcal{Q} = \dd U - \delta W. 
\end{align}
If the only non-zero contribution to the internal energy comes from the electrostatic interaction, 
we have 
\begin{align}
	\dd U &=\dd\!\left(\frac{\varepsilon_0\varepsilon_r A}{2} 
    \int\limits_0^L (E(z))^2\dd z\right)\\
	&= \varepsilon_0\varepsilon_r A \int\limits_0^L \left(\dd E(z) E(z)\right)\dd z\\
	&= eA \int\limits_0^L\left(\dd q(z)\  \psi(z)\right)\dd z\,.
\end{align}
In the last step, an integration by parts is performed. Further, one should keep in mind 
that the charge distribution $q(z)$ contains contributions from both the ions 
$\sum_{i}z_i \rho_i(z)$ and the surface 
charges on the electrodes $\sigma$. 

For a process of duration $\mathcal{T}$, the change in internal energy follows as
\begin{align}
	\Delta U &= \int\limits_0^\mathcal{T}\frac{\dd U}{\dd t}\dd t\\
	&=eA\int\limits_0^\mathcal{T}\int\limits_0^L \frac{\dd q(z)}{\dd t}\psi(z)\dd z\ \dd t\label{equ:energy_current_I}\\
	&=-A\int\limits_0^\mathcal{T}\int\limits_0^L I(z) E(z)\dd z\ \dd t,\label{equ:energy_current_II}
\end{align}
where the continuity equation is used and another integration by parts is performed. 

Similarly, the electric work during (dis)charging can be written as 
\begin{align}
	\Delta W_\text{el} &= \int\limits_0^\mathcal{T} \frac{\dd Q}{\dd t} \Delta\Psi \dd t\,.\label{equ:work_current}
\end{align}
Subtracting \cref{equ:work_current} from \cref{equ:energy_current_II} (or 
respectively \cref{equ:energy_current_I}), one is left with the ionic currents. 
However, one should keep in mind that this difference yields the corresponding heat only 
if there are no other work terms. In a grand canonical charging process, for example, 
one would also get a work term due to a particle flux into/out of the system 
\begin{align}
	\Delta W_\text{ch} &= \sum\limits_i \mu_i\Delta N_i\,.
\end{align}
If one wants to calculate the reversible heat, an infinitely slow (dis)charging process 
must be regarded where the system is in equilibrium at every time. 
Variable substitutions 
in \cref{equ:energy_current_I,equ:work_current}, 
replacing the time integrals by integrals over surface charge density, yield 
\begin{align}
	\mathcal{Q}_\text{rev}^{\text{II}} &= A \int\limits_0^{\sigma}\int\limits_0^L \psi(z) 
  \frac{\dd}{\dd \sigma'}\left(\sum\limits_{i}z_i \rho_i(z)\right)\dd z\ \dd \sigma'\,.
\end{align}
Note that we introduced the index II solely for conformity with the main text. 
This equation is very useful to calculate the reversible heat from DFT data. 

\section{Derivation of $\mathcal{Q}_\text{rev}^{\text{I}}$}\label{appendix:QrevI}

For the \textsc{Gouy-Chapman} solution, the heat flow near a single charged wall follows from 
\cref{equ:GC_surface_charge_density,eq:QrevI'} and the definition of $\bar{\sigma}$ as 
\begin{widetext}
\begin{align}
	\frac{\mathcal{Q}_\text{rev}^{\text{I},\text{GC}}}{A\kB T}&=-2\int\limits_0^{\sigma} 
    \left[\frac{\partial T\sinh^{-1}\left(\sigma'/\bar{\sigma}\right)}{\partial T} \right]_{\alpha, \sigma'}\,\dd\sigma'\\
	&=-2\int\limits_0^{\sigma} \left[\sinh^{-1}\left(\frac{\sigma'}{\bar{\sigma}}\right) - 
    \frac{ \sigma'}{\sqrt{\sigma'^2+\bar{\sigma}^2}}\left(\frac{\partial\ln \bar{\sigma}}{\partial \ln T}\right)_{\alpha}  \right]\,\dd\sigma'\\
	&=- \bar{\sigma}\left\{\Phi\sinh\left(\frac{\Phi}{2}\right) +2\left( 1- \cosh\left(\frac{\Phi}{2}\right) \right)\left[1+\frac{1}{2}  \frac{\partial \ln [\varepsilon_r(T) T]}{\partial \ln T} + \frac{1}{2}  \left(\frac{\partial \ln \bar{\rho}_+}{\partial \ln T} \right)_{\alpha}\right]\right\} . 
	\label{eq:Qrevexplicit}
\end{align}
\end{widetext}	
Here, $\alpha$ stands for the variable(s) kept fixed during the partial $T$-derivative.
If we consider the ensemble of fixed ionic concentration ($\alpha=\bar{\rho}_{+}=\bar{\rho}_-$), 
the last term \cref{eq:Qrevexplicit} drops out. 
If in addition $\partial(\varepsilon_r(T)T)/\partial T=0$, which is accurate for water (cf. p. 69 in Ref.~\cite{THEODOOR199061}), \cref{eq:Qrevexplicit} yields
\begin{align}
		\frac{\mathcal{Q}_\text{rev}^{\text{I},\text{GC}}}{A\kB T}&=-\bar{\sigma}\left[2-2\cosh\left(\frac{\Phi}{2}\right)+\Phi\sinh\left(\frac{\Phi}{2}\right)\right]\label{eq:QrevSimple}\\
	&=-\left(\Delta\Omega_{\text{ent}}^{\text{GC}} + \Delta\Omega_{\rm el}^{\text{GC}}\right)\,,
\end{align}
the same as Eq.~(S11) in Ref.~\cite{janssen_prl119_2017} (up to a typo in their $\Omega$ subscript).
This, however, would mean that at every point of charging 
the amount of electric work put into the system would flow out of the system in the form 
of reversible heat. Thus, the internal energy of the system would remain constant during 
charging, which would be surprising. This may be resolved by including some explicit model for 
the solvent responsible for the $T$ dependence of $\varepsilon_r(T)$. The explicit model would yield 
further entropic contributions and may also resolve the problem that for $T$-dependent 
$\varepsilon_r(T)$ one gets $\mathcal{Q}_\text{rev}^{\text{I}}\neq\mathcal{Q}_\text{rev}^{\text{II}}$, 
as was found in Ref.~\cite{janssen_prl119_2017}.
If $\partial(\varepsilon_r(T))/\partial T=0$ is considered instead, we find	
\begin{align}
	\frac{\mathcal{Q}_\text{rev}^{\text{I},\text{GC}}}{A\kB T}	
	&=-\bar{\sigma}\left[3-3\cosh\left(\frac{\Phi}{2}\right)+\Phi\sinh\left(\frac{\Phi}{2}\right)\right]\\
	&=-\Delta\Omega_{\text{ent}}^{\text{GC}} . 
\end{align}

In the grand canonical ensemble ($\alpha=\mu_+=\mu_-$), we make use of 
$\mu_\pm = \kB T \ln \left(\Lambda_{\pm}^3 \bar{\rho}_\pm\right)$ to find 
\begin{align}
 \left(\frac{\partial \ln \bar{\rho}_\pm}{\partial \ln T} \right)_{\mu_\pm} 
&  = \frac{3}{2}-\frac{\mu_\pm}{\kB T} \,. 
\end{align}
For the case that $\partial[\varepsilon_r(T)]/\partial T=0$, one now finds 
\begin{align}
		\frac{\mathcal{Q}_\text{rev}^{\text{I},\text{GC}}}{A\kB T} &= 
      -\bar{\sigma}\left\{\Phi\sinh\left(\frac{\Phi}{2}\right) 
      +\left( 1- \cosh\left(\frac{\Phi}{2}\right) \right)\left[\frac{9}{2}-\frac{\mu_+}{\kB T}\right]\right\}\,,
\end{align}
which is \cref{eq:GCgc} of the main text.

\section{Adsorption in the canonical \textsc{Gouy-Chapman} solution}\label{appendix:canonical_GC}

In a canonical charging process, the total number of ions must be conserved. To derive the 
corresponding equation for $\bar{\rho}_\pm$, a system free of overlap is assumed. For 
simplicity, every electrode is assumed to be one charged hard wall in order to get rid of an 
additional factor of $2$ that cancel anyway for the ratio $-\mathcal{Q}_\text{rev}/W_\text{el}$. 

Following Ref.~\cite{boon2011blue} (Eqs. 7-10), we see that the total number of ions per species in our system can be written as
\begin{align}
	\frac{N_\pm}{A} &= \frac{\Delta N}{A} + \bar{\rho}_\pm L\,,
\end{align}
where $\Delta N$ is
\begin{align}
	\frac{\Delta N}{A} &= \bar{\sigma}\left[\cosh\left(\frac{\Phi}{2}\right)-1\right].\label{equ:particle_number_change}
\end{align}
This equation can be solved for $\bar{\rho}_\pm$ where one should keep in mind that 
$\bar{\sigma}$ is $\bar{\rho}_\pm$ dependent. 

In combination with \cref{equ:GC_surface_charge_density} we have thus two equations to numerically 
search for combinations of $\sigma$, $\Phi$, and $\bar{\rho}_\pm$ that solve these equations. 
From $\bar{\rho}_\pm$ one can simply calculate $\mu_i$ and thus we can calculate $W_\text{el}$ 
and $F$ then.
\end{appendix}

\end{document}